\documentclass[a4paper,10pt,twocolumn]{article}

\usepackage{graphicx}
\usepackage[a4paper,lmargin=1.6cm,rmargin=1.6cm,tmargin=2.5cm,bmargin=2.5cm]{geometry}
\usepackage{authblk}
\usepackage{titlesec}
\usepackage{url}
\usepackage{caption}
\usepackage{fancyhdr}

\newcommand{\citep}[1]{\cite{#1}}
\newcommand{\citet}[1]{\cite{#1}}
\newcommand{\zeroindexLarge}{0}
\newcommand{\zeroindexSmall}{{\rm o}}
\newcommand{\Vzero}{V_\zeroindexLarge}
\newcommand{\nzero}{n_\zeroindexSmall}
\newcommand{\vzero}{v_\zeroindexSmall}
\newcommand{\epsilonzero}{\epsilon_\zeroindexSmall}

\title{\Large{\textbf{CUBESAT TESTING OF COULOMB DRAG PROPULSION}}}

\titleformat{\section}{\bfseries\uppercase}{\thesection}{1em}{}
\titleformat{\subsection}{\bfseries}{\thesubsection}{1em}{}
\author[1]{Pekka Janhunen}
\author[1]{Jouni Envall}
\author[1]{Petri Toivanen}
\author[2]{Timo Rauhala}
\author[2]{Edward H{\ae}ggstr\"om}
\author[3]{Tor-Arne Gr\"onland}

\affil[1]{Finnish Meteorological Institute, POB--503, FI--00101,
  Helsinki, Finland}
\affil[2]{University of Helsinki, Electronics Research Laboratory, Finland}
\date{}
\affil[3]{Nanospace AB, Uppsala, Sweden}

\begin{document}

\maketitle
\thispagestyle{fancy}

\begin{abstract}
In Coulomb drag propulsion, a long high voltage tether or system of
tethers gathers momentum from a natural plasma stream such as solar
wind or ionospheric plasma ram flow. A positively polarised tether in
the solar wind can be used for efficient general-purpose
interplanetary propellantless propulsion (the electric solar wind sail
or E-sail), whereas a negatively polarised tether in LEO can be used
for efficient deorbiting of satellites (the plasma brake). Aalto-1 is
a 3-U cubesat to be launched in May 2016. The satellite carries three
scientific experiments including 100 m long Coulomb drag tether
experiment. The tether is made of four 25 and 50 micrometre diameter
aluminium wires that are ultrasonically bonded together every few
centimetre intervals. The tether can be charged by an onboard voltage
source up to one kilovolt positive and negative.  The Coulomb drag is
measured by monitoring the spin rate.
\end{abstract}

\section{Introduction}

\begin{figure}
\includegraphics[width=\columnwidth]{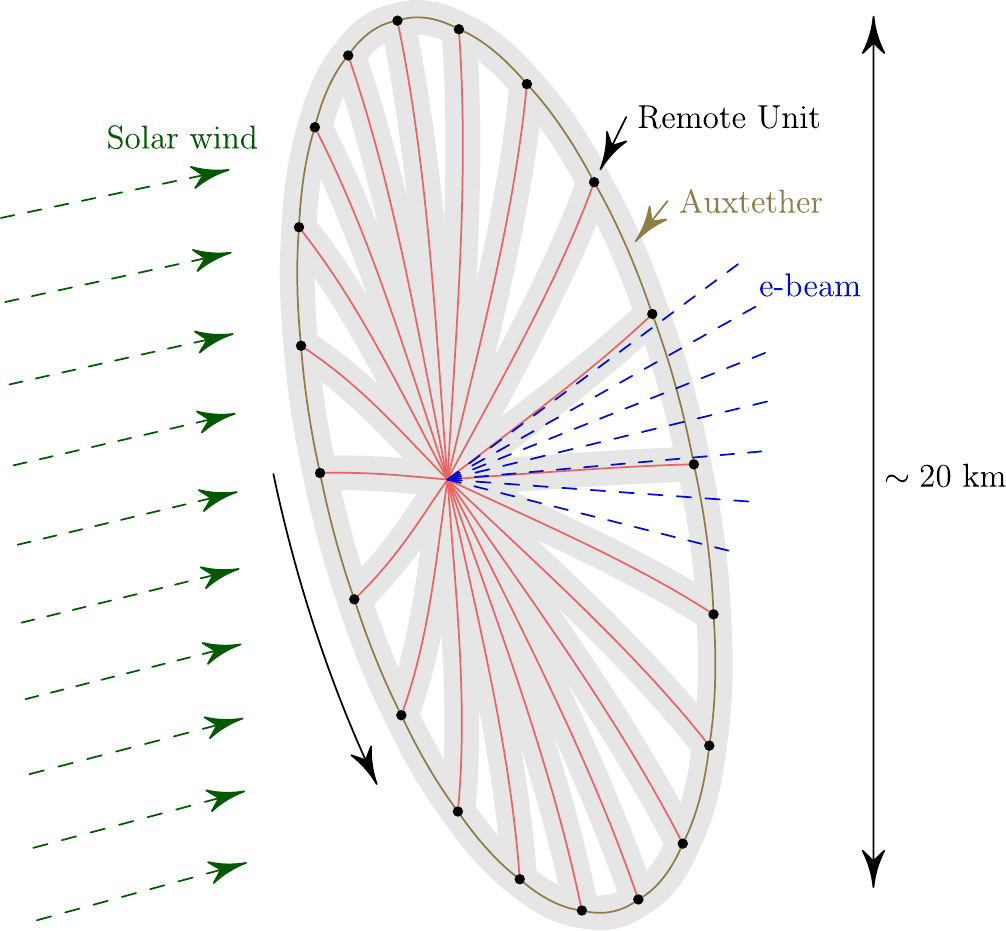}
\caption{
Centrifugally stabilised E-sail with electric auxiliary tethers to
enable spin control during flight \citep{paper39}. 
}
\label{fig:Esail3D}
\end{figure}

Coulomb drag propulsion means putting a long, thin and charged tether
into a natural plasma flow and tapping momentum from the flow by
Coulombic deflection of the flow ions by the electric field that
surrounds the tether. Coulomb drag propulsion was first proposed
\citep{paper1,paper2} for interplanetary propulsion by the electric
solar wind sail (electric sail, E-sail), Fig.~\ref{fig:Esail3D}.

The E-sail has potentially revolutionary performance level in
comparison to other propulsion systems \citep{paper9}. The tether
weighs only 10 grams per kilometre and produces a thrust of $\sim 0.5$
mN/km at 1 au distance. The E-sail thrust scales as proportional to
$1/r$ where $r$ is the solar distance. The reason is that while the
solar wind dynamic pressure decays as $\sim 1/r^2$, the plasma Debye
length (by which the electric field penetration distance and hence the
virtual sail size scales) varies as $\sim r$, thus giving an overall
$1/r$ dependence for the thrust. For example, hundred 20 km long
tethers would weigh 20 kg and they would produce 1 N thrust at 1 au
which gives a 30 km/s velocity change pear year for a 1000 kg
spacecraft.

E-sail thrust magnitude can be easily controlled between zero and a
maximum value by changing the voltage of the tethers. The tether
voltage is maintained by continuously operating an electron gun which
pumps out negative charge from the system, hence tether voltage can be
actuated easily by changing the current and voltage of the electron
gun beam. The power consumption of the electron gun is moderate (700 W
nominally at 1 au for large 1 N sail) and it scales as $1/r^2$,
i.e.~in the same way as the illumination power of solar panels. The
power consumption stems from the electron current gathered from the
surrounding solar wind plasma by the positively charged tethers which
can be estimated by so-called orbital motion limited (OML) cylindrical
Langmuir probe theory \cite{paper2},
\begin{equation}
\frac{d I}{d z} = e \nzero \sqrt{\frac{2 e \Vzero}{m_e}} \left(2 r_w^{\rm base} +
4.5\times r_w^{\rm loop}\right).
\label{eq:dIdz}
\end{equation}
Here $e$ is the electron charge, $\nzero$ is the solar wind plasma
density, $\Vzero$ is the tether bias voltage, $m_e$ is electron mass,
$r_w^{\rm base} = 25\,\mu$m is the base wire radius of the
micrometeoroid-resistant tether, which is assumed to have four wires
in Eq.~(\ref{eq:dIdz}) with parallel wire radius
$r_w^\mathrm{base}=25\,\mu$m and loop wire radius
$r_w^\mathrm{loop}=12.5\,\mu$m.

\begin{figure}
\includegraphics[width=8.5cm]{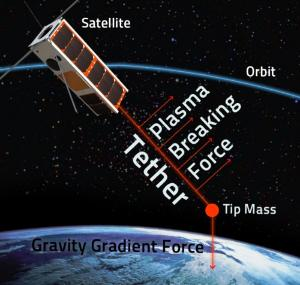}
\caption{
Negatively biased gravity-stabilised Coulomb drag plasma brake in LEO
for satellite deorbiting \citep{KestilaEtAl2013}.
}
\label{fig:PlasmaBrake}
\end{figure}

Later it was found \citep{paper3,plasmabrake} that Coulomb drag
propulsion can also be used for satellite deorbiting in low Earth
orbit (LEO). Contrary to the solar wind, in LEO it seems more
attractive to use negative tether polarity because it needs less power
than a positive polarity one and because the balancing ion emitter can
in LEO conditions often be implemented simply by the satellite's
conducting surface. The LEO application is called the plasma brake
(Fig.~\ref{fig:PlasmaBrake}). According to simulations, a 5 km long negatively
charged plasma brake tether weighing 0.055 kg could produce 0.43 mN
braking force which is enough to reduce the orbital altitude of a 260
kg object by 100 km per year \citep{paper24}.

Part of the material in this paper is taken from the Space Propulsion
2014 proceedings paper \citep{Koln2014}.

\section{Physics of Coulomb drag}

\begin{figure*}[ht]
\centering
\includegraphics[width=0.9\textwidth]{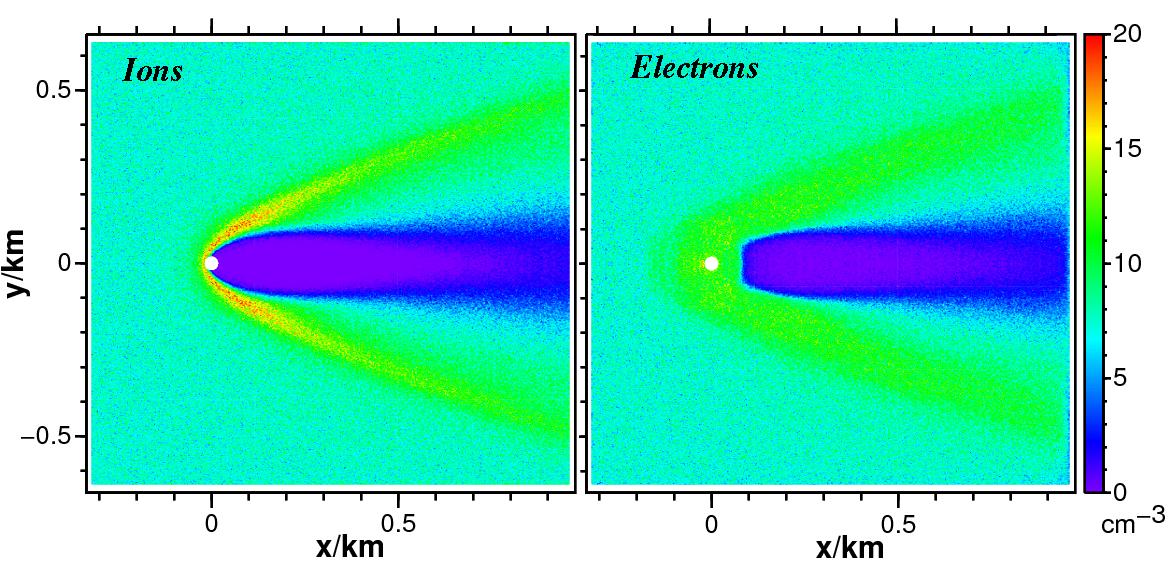}
\caption{
PIC simulation of positively charged E-sail tether interacting with streaming
solar wind plasma \citep{ASTRONUM2011}. Solar wind (density 7.3 cm$^{-3}$, speed
400 km/s) arrives from the left and interacts with +5.6 kV charged
tether at origin (white dot).
}
\label{fig:PositiveTether}
\end{figure*}

\begin{figure*}[ht]
\centering
\includegraphics[width=\columnwidth]{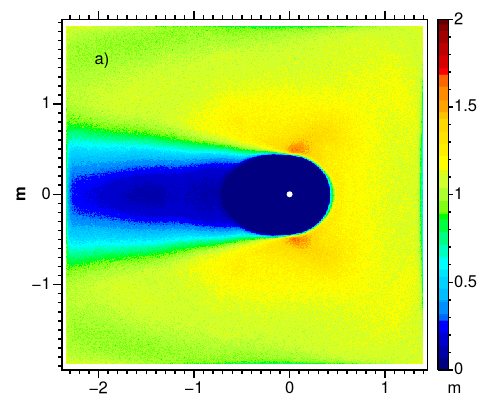}\nobreak
\includegraphics[width=\columnwidth]{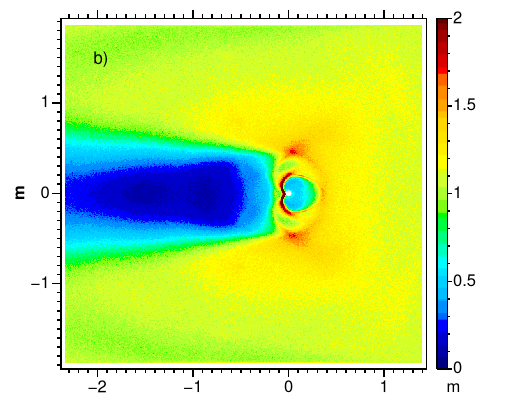}
\caption{PIC simulation of negative tether in LEO
  \citep{paper24}. Electron density (a) and density (b), normalised
  to ambient plasma stream density. Flow arrives from the right and
  tether voltage is -337 V.
}
\label{fig:NegativeTether}
\end{figure*}

When plasma streams past a charged thin tether, the tether's electric
field penetrates some distance into the plasma and deflects the
charged particles of the stream. Because electrons are lightweight,
the momentum flux carried by them is negligible so it is enough to
consider the deflection of ions. Both positively and negatively biased
tethers cause ion deflection and hence Coulomb drag. A positive tether
deflects positively charged ions by repelling them
(Fig.~\ref{fig:PositiveTether}). A negative tether deflects ions by
attracting them so that their paths cross behind the tether
(Fig.~\ref{fig:NegativeTether}).

\subsection{Positively biased tether}

A positively biased tether repels stream ions and attracts
electrons. When the potential is turned on, a population of trapped
electrons gets formed \cite{paper2}. In most of the literature
concerning biased tethers, it is implicitly or explicitly assumed that
trapped electrons are not present in the asymptotic state. In a
multi-tether starfish-shaped E-sail geometry (Fig.~\ref{fig:Esail3D}),
trapped electron orbits are chaotised whenever the electron visits the
central ``hub'' which is the spacecraft, and chaotised electrons have
a small nonzero probability of getting injected into an orbit which
takes them to collision course with a tether wire so that trapped
electrons are removed by this mechanism in few minute timescale in
nominal 1 au solar wind \cite{paper6}. It might be that other
processes such as plasma waves occur in nature which speed up the
process. By PIC simulations alone it is not easy to predict how many
trapped electrons are present in the final state, although the
question was recently analysed also using a novel, indirect approach
\citep{paper35}.

The E-sail thrust per tether length $dF/dz$ is given by
\begin{equation}
\frac{dF}{dz} = K P_\mathrm{dyn} r_s
\label{eq:dFdz}
\end{equation}
where $K$ is a numerical coefficient of order unity, $P_\mathrm{dyn} =
\rho v^2$ is the dynamic pressure of the plasma flow and $r_s$ is the
radius of the electron sheath (the penetration distance of the
electric field into the plasma). The quantity $r_s$ can be inferred
from recent laboratory measurements of Siguier et
al.~\cite{SiguierEtAl2013} where Ar$^{+}$ plasma (ion mass $m_i=40$ u) of
density $\nzero=2.4\cdot 10^{11}$ m$^{-3}$ accelerated to 20 eV bulk flow
energy (hence speed $v=9.8$ km/s) was used and let to interact with $r_w=2.5$ mm
radius metal tether biased to $\Vzero=100$ V and 400 V in two experiments. At
$\Vzero=100$ V the sheath radius as visually determined from their Figure
7 is $r_s=12$ cm and at 400 V it is $r_s=28$ cm (from their Figure 8). For estimating
the corresponding $dF/dz$, let us assume $K=2$ in the above formula
[Eq.~(\ref{eq:dFdz})]. This corresponds to assuming that ions incident
on the sheath are on average deflected by 90$^{\circ}$ (notice that
the size of the virtual obstacle made by the sheath is twice its
radius). We think that this is a reasonable first estimate since ions
arriving head-on towards the tether are reflected backwards while ions
arriving near the boundaries of the sheath are probably deflected
less than 90$^{\circ}$. In their experiment $P_\mathrm{dyn} = 1.54$ $\mu$Pa so
Eq.~\ref{eq:dFdz} gives $dF/dz=370$ nN/m and $dF/dz=860$ nN/m for
$\Vzero$ equal to 100 V and 400 V, respectively.

Let us compare these experimentally inferred values with theoretical
estimates. A simple theoretical estimate for the sheath radius is the
effective Debye length
\begin{equation}
\lambda_D^{\rm eff} = \sqrt{\frac{\epsilon_0
    \left(\Vzero-V_1\right)}{e \nzero}}
\label{eq:lambdaDeff}
\end{equation}
where $V_1=(1/2)m_i v^2/e$ is the stream ion bulk flow energy. The
expression (\ref{eq:lambdaDeff}) for the effective Debye length is
obtained from the usual formula for ordinary electron Debye length by
replacing the electron temperature by the tether voltage. We also
subtract the bulk energy term $V_1$ to model the fact that if the
tether voltage is lower than the bulk energy, it can no longer reflect
back or stop ions but only weakly deflects them even if they arrive
with zero boost parameter; the subtraction of $V_1$ however has only
modest impact to our results. If one takes $r_s$ to be equal to
$\lambda_D^{\rm eff}$ in Eq.~(\ref{eq:dFdz}), one obtains $dF/dz$
equal to 420 nN/m and 910 nN/m for $\Vzero$ equal to 100 V and 400 V,
respectively.

Theoretical E-sail thrust formulas of \cite{paper9} contain the average
electron density $n_e$ inside the sheath as a free parameter, the
choice $n_e=0$ giving the largest E-sail thrust. Assuming $n_e=0$ and
applying the formulas for the experimental parameters of Siguier et
al.~\citep{SiguierEtAl2013}, one obtains 220 nN/m and 740 nN/m thrust
per length for $\Vzero$ equal to 100 V and 400 V, respectively.

We summarise the experimental and theoretical results in Table
\ref{tab:numeric-comparison}.

\begin{table}
\caption{Comparison of experimental and theoretical E-sail thrust per length in LEO-like conditions.}
\label{tab:numeric-comparison}
\begin{tabular}{lll}
\hline
                                               & $\Vzero$=100 V & $\Vzero$=400 V\vphantom{\AA}  \\
\hline
Siguier\vphantom{\AA} et al.~\citep{SiguierEtAl2013}         & 370 nN/m       & 860 nN/m        \\
$\lambda_D^{\rm eff}$, Eq.~\ref{eq:lambdaDeff} & 420 nN/m       & 910 nN/m        \\
Theory of \citep{paper9} /w $n_e$=0            & 220 nN/m       & 740 nN/m        \\
\hline
\end{tabular}
\end{table}

We conclude from Table \ref{tab:numeric-comparison} that experimental
results of \citep{SiguierEtAl2013} are consistent with the assumption
of no trapped electrons, i.e.~maximal E-sail thrust.

\subsection{Negatively biased tether}

\label{subsect:negbias}

In the negative polarity case, electrons are simply repelled by the
tether (Fig.~\ref{fig:NegativeTether}) and hence the physics of
electrons is simple. We believe that PIC simulations
therefore have a good chance of predicting the thrust correctly in the negative polarity case. Using a new
supercomputer, a comprehensive set of negative polarity PIC
simulations for LEO-like parameters was recently conducted
\citep{paper24} and it was found that the following formula gives a
good fit to the PIC simulations:
\begin{equation}
\frac{dF}{dz} =
3.864 \times P_\mathrm{dyn} \sqrt{\frac{\epsilonzero \tilde{V}}{e \nzero}}
\exp\left(-V_i/\tilde{V}\right).
\label{eq:dFdzneg}
\end{equation}
Here $\vzero$ is the ionospheric plasma ram flow speed relative to spacecraft (assumed to be perpendicular to
the tether or else $\vzero$ denotes only the perpendicular component), $P_\mathrm{dyn} = m_i \nzero \vzero^2$ is the flow dynamic pressure,
$m_i$ is the ion mass (typically the plasma is singly ionised atomic
oxygen so that $m_i\approx 16$ u),
\begin{equation}
\tilde{V} = \frac{\vert\Vzero\vert}{\ln(\lambda_D^{\rm eff}/r_w^{*})},
\end{equation}
$r_w^{*}$ is the tether's effective electric radius (Appendix
  A of \citep{paper2}), $\lambda_D^{\rm eff}=\sqrt{\epsilonzero \vert\Vzero\vert/(e \nzero)}$
is the effective Debye length and $V_i=(1/2)m_i \vzero^2/e$ is the
bulk ion flow energy in voltage units. The effective electric radius
is approximately given by $r_w^{*} = \sqrt{b r_w} \approx 1$ mm, where $r_w$ is the
tether wire radius, typically 12.5-25 $\mu$m, and $b$ is the tether
width, typically 2 cm (a rough value of $b$ is sufficient to
  know because $r_w^*$ enters into
  Eq.~(\ref{eq:dFdzneg}) only logarithmically).

Thus, although experimental confirmation is needed, there is good
reason to believe that Eq.~\ref{eq:dFdzneg} describes LEO plasma brake
thrust well. The only exception is that if the geomagnetic field
is predominantly oriented along the tether, the interaction becomes
turbulent and the thrust is moderately reduced \cite{paper24}. The
reduction grows with increasing voltage: it is 17\% at $-320$ V bias
voltage and 27\% at $-760$ V. For a vertical gravity-stabilised plasma
brake tether in polar orbit, efficiency reduction with respect to
Eq.~(\ref{eq:dFdzneg}) is thus expected at high latitudes.

We emphasise that Eq.~(\ref{eq:dFdzneg}) has thus far only been
verified with simulations in LEO plasma environment conditions. If
negative polarity Coulomb drag devices would become relevant in the
future also in other plasma conditions, the applicability of
Eq.~(\ref{eq:dFdzneg}) should be considered carefully on a case by
case basis.

\section{ESTCube-1 experiment}

ESTCube-1 was a 1-U Estonian cubesat which flew in 2013-2015
\citep{LattEtAl2014}. ESTCube-1 carried a preliminary 10 m version of
our tether experiment \citet{EnvallEtAl2014}. The experiment did not
work because the tether reel failed to rotate. A reason was identified
after the launch in ground testing: a gold-plated slipring was pressed by a
steel spring contactor and during launch vibration, the contactor
carved a small hole into the slipring which was enough to prevent the
piezoelectric motor from turning the reel. Although the error was in
our tether payload, it was a ``shallow'' problem whose reason was
insufficient vibration testing before launch, which in turn was due to
5-month accelerated schedule of the satellite in a late phase, due to
sudden emergence of a free launch opportunity and the satellite team's
decision to accept it.  The error was thus not directly related to any
deeper aspects of Coulomb drag tether technology.

\section{Aalto-1 satellite}

\begin{figure}[h]
\centering
\includegraphics[width=0.85\columnwidth]{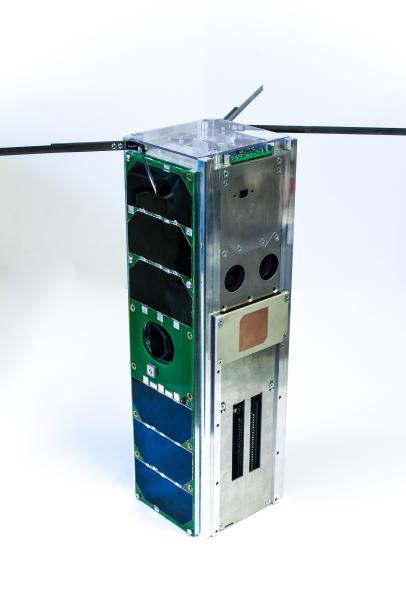}
\caption{Aalto-1 satellite. Photo: Tuomas Tikka, Aalto University.}
\label{fig:A-1}
\end{figure}

Aalto-1 \citep{KestilaEtAl2013} (Fig.~\ref{fig:A-1}) is a Finnish 3-U,
4 kg cubesat which will be launched in polar LEO in May 2016 onboard
SpaceX Falcon-9 rocket. Like ESTCube-1, the Aalto-1 cubesat is built
by university students while the payloads are made by research
groups. Aalto-1 carries three payloads, AASI imaging Fabry-Perot
spectrometer (hyperspectral camera) by VTT, a compact radiation
monitor (RADMON) by the University of Turku and our Coulomb drag
tether experiment.

\begin{figure*}[ht]
\centering
\includegraphics[width=0.9\columnwidth]{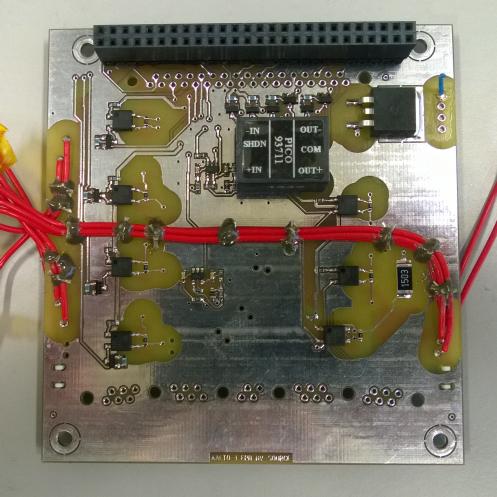}\nobreak\quad \nobreak
\includegraphics[width=0.9\columnwidth]{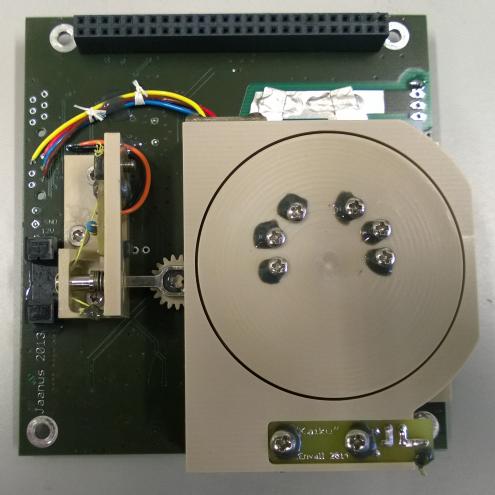}
\caption{
Aalto-1 tether payload HV card (left) and motor card (right).
}
\label{fig:A1-cards}
\end{figure*}

\section{Aalto-1 tether experiment}

The tether experiment of Aalto-1 is a debugged version of the
ESTCube-1 experiment: known problems have been fixed, testing was more
complete than in case of ESTCube-1, and more diagnostics were added so
that if something goes wrong, we will be able to know more precisely
what it was.

The length of the tether in Aalto-1 is 100 m i.e.~ten times longer
than in ESTCube-1. The tether is made of four 25 and 50 micrometre
diameter aluminium wires that are ultrasonically bonded together every
few centimetre intervals
\citep{SeppanenEtAl2011,SeppanenEtAl2013}. The tether can be charged
by an onboard voltage source up to one kilovolt positive or negative
with respect to surrounding plasma. The negative polarity Coulomb drag
experiment of Aalto-1, in particular, is expected to yield a practical
demonstration of plasma braking by significantly lowering the orbit of
the satellite or even deorbiting it entirely.

Figure \ref{fig:A1-cards} shows the high-voltage (HV) and motor cards
of the Aalto-1 tether payload. The motor card also hosts the tether
reel. The lateral size of both cards is about $10\times10$ cm (a
standard cubesat circuit board) and together they take 50.5\,mm of
vertical space so that both cards together take 0.5U of volume. The
motor card weighs 148\,g and the HV card weighs 114\,g so that together
they weigh 262\,g.

When the tether experiment is started, the satellite is spun up using
magnetic coils, the 1.2\,g endmass and reel locks are released and the
motor is commanded to turn the reel. The tether then starts to deploy
and the centrifugal pull of the endmass keeps it under proper
tension. When deployment proceeds, the tether tension first increases
almost linearly with tether length. When the moment of
inertia of the deployed part of the tether has become comparable to
the satellite's moment of inertia, the tension reaches a maximum at
$\sim 1.5$\,m tether length and then starts to decrease again because
the satellite's rotation slows down due to angular momentum
conservation.

We deploy first about 10\,m of the tether and then start positive
tether polarity experiments by operating the satellite's cold cathode
electron gun \citep{KleshchEtAl2015}. The reason for starting positive
mode experiments with relatively short tether is to ensure that the
current produced by the gun is sufficient to balance the electron
current gathered by the positively biased tether from the
plasma. Otherwise there would be a risk that the tether voltage would
drop an unknown amount below the electron gun beam voltage so that the
tether voltage would become unknown.

The tether spins in equatorial plane and we perform the Coulomb drag
experiment primarily over high latitudes so that the plasma ram flow
is in the plane of the tether. We measure the Coulomb drag thrust in
the following indirect way. We keep the tether voltage on whenever the
tether is moving downstream with respect to plasma ram flow and keep
it off when it moves upstream. In this way the Coulomb drag that is
exerted on the tether accelerates the tether's and the satellite's spin
during each spin period. Because the effect accumulates over
successive spins, it relatively quickly should yield a change in the
satellite's spinrate which is large enough to be easily measurable by
the attitude control system and also by simple and robust methods such
as looking at the periodicity of the sun sensor signals or power
produced by a particular solar panel. The purpose of positive mode
experiments is to determine the positive mode thrust as a function of
plasma density, which can be estimated based on orbital location and
geophysical conditions from ionospheric models such as International
Reference Ionosphere (IRI) \citep{BilitzaEtAl2011}.

After successful positive mode tests at $\sim 10\,$m length, a similar
series of negative mode experiments is also carried out. The negative
mode does not need the electron gun, but only a HV source which forces
the tether negative with respect the satellite frame. Conducting parts
of the satellite surface then go weakly positive until they gather
enough thermal electron flux from the plasma to balance the negative tether's
gathered plasma ion current. As a result, the tether's negative
voltage will be very nearly the same as the voltage given by the
source so that the tether will be in a known negative voltage.

After successful experiments at $\sim 10\,$m length, one of the
Coulomb drag effects is used to increase the spinrate so that tether
deployment can continue beyond 10\,m while keeping a tether
tension which is large enough to keep it straight and to overcome any
Coulomb drag effect by a large margin. The process is continued until
all 100 m have been deployed, likely interleaving deployment phases
and spin acceleration phases 2-3 times. At the end, the voltage is
left on the tether all the time in negative mode so that the spinrate
is no longer modified, but the satellite's centre of mass experiences
the drag and the orbit is lowered. If the experiment can run for
months in this mode, clear lowering in satellite's orbital altitude
should be observable, of order 5-10 km per month. In the very best
case, if the satellite stays alive for a long enough time, it's
possible that full deorbiting will eventually result due to the plasma
brake effect followed by engaging the atmospheric drag at lower
altitudes.

\section{Expected plasma brake thrust}

Recall from section \ref{subsect:negbias} that for LEO parameters,
Eq.~\ref{eq:dFdzneg} is a good fit to PIC simulations (except for
modest thrust reduction due to turbulence when the dominant component
of the geomagnetic field is along the tether) which in turn are
expectedly good models of reality in case of negative tether voltage.

One noteworthy fact is that LEO plasma brake thrust according to
Eq.~\ref{eq:dFdzneg} is proportional (through linear dependence on
$P_\mathrm{dyn}$) to the ion mass $m_i$. Thus, plasma brake thrust is 16
times larger in pure oxygen O$^{\circ}$ plasma than in pure proton
plasma.

For example, at $\Vzero=-1$ kV, $\nzero=3\cdot 10^{10}$ m$^{-3}$,
$v=7.5$ km/s and $m_i=16$ u, Eq.~\ref{eq:dFdzneg} gives 85 nN/m
thrust. In the negative bias case, usable tether voltage is limited by
onset of electron field emission. We think that above 1-2 kV, field
emission might start to become an issue.

\begin{table}[h]
\caption{Predicted plasma brake thrust (nN/m) for solar min/solar max
  conditions.}
\label{tab:plasmabrakethrust}
\begin{tabular}{llll}
\hline
Altitude\vphantom{\AA} & MLT 12-00 & MLT 06-18  & Average \\
\hline
& & & \\[-10pt]
700 km\vphantom{\AA} & 47/157    & 42/140     & 44/149  \\
800 km & 33/117    & 30/108     & 32/112  \\
900 km & 25/88     & 22/84      & 23/86   \\
\hline
\end{tabular}
\end{table}

Table \ref{tab:plasmabrakethrust} gives plasma brake thrust based on
Eq.~\ref{eq:dFdzneg}, assuming $\Vzero=-1$ kV, $v=7.5$ km/s and using
plasma density and chemical composition taken from the IRI-2012
ionospheric model \citep{BilitzaEtAl2011}, for noon-midnight (mean local time MLT 12-00) and
dawn-dusk (MLT 06-18) polar orbits and for solar minimum and maximum
ionospheric conditions. We see from Table \ref{tab:plasmabrakethrust}
that the dependence on solar cycle is relatively significant, about
factor 3.5. The solar cycle dependence is due to increased plasma
density and increased oxygen abundance during solar maximum
conditions. There is obviously also an altitude dependence. Below 700
km the thrust would continue to increase until $\sim 400-500$ km,
provided that the hardware is designed to take advantage of it. The
dependence on MLT is weak.

As a numerical example, consider a 10 km long plasma brake tether which
starts bringing down a debris object of 200 kg mass from 800 km
circular orbit in an MLT which is average between dawn-dusk and
noon-midnight. The required $\Delta v$ from 800 km to 700 km is 53.5
m/s and from 700 km to 400 km 165 m/s. During solar minimum,
deorbiting from 800 km to 700 km takes 0.88 years and the rest from
700 km to 400 km (assuming the same thrust as at 700 km) takes 2.4
years, thus altogether 3.25 years. During solar maximum the
800$\to$700 km deorbiting takes 0.25 years and 700$\to$400 km 0.7
years, thus altogether 0.96 years. These estimates are conservative
since in reality plasma density and oxygen concentration and hence
plasma brake thrust continue to grow below 700 km.

\section{Discussion and conclusions}

Based on plasma simulations and an indirect laboratory result, there
is good reason to think that the magnitude of tether Coulomb drag is
in very useful range at least for (1) positive tether in solar wind
(E-sail) and (2) negative tether in LEO (plasma brake).

Coulomb drag propulsion in both positive and negative mode will soon
be attempted to be measured by Aalto-1 satellite. Aalto-1 is a 3-U
cubesat to be launched in May 2016 in polar LEO. Ideally we will get a
measurement of positive and negative mode Coulomb drag as function of
plasma density, tether length and voltage. The results can then be
compared with earlier simulation predictions. We should also be able
to observe measurable lowering of the satellite orbit using the
negative mode at 100 m tether length. In the very best case, if the
satellite and the experiment stay alive long enough, we could even
observe reentry into atmosphere.

\emph{Acknowledgements.} This research was partially financed within
the European Community's Seventh Framework Programme ([FP7/2007-2013])
under grant agreement number 262733, the Academy of Finland grant
250591 and European Space Agency grant 4000115856/15/NL/PS/gp.


\begin{thebibliography}{00}

\bibitem{paper1}
Janhunen, P., Electric sail for spacecraft propulsion, \emph{J.~Prop.~Power,}
\textbf{20}, 763--764, 2004.

\bibitem{paper2}
Janhunen, P.~and A.~Sandroos, Simulation study of solar wind push on
a charged wire: basis of solar wind electric sail propulsion,
\emph{Ann.~Geophys.,} \textbf{25}, 755--767, 2007.

\bibitem{paper39}
Janhunen, P.~and P.~Toivanen, TI tether rig for solving secular
spinrate change problem of electric sail,
\url{http://arxiv.org/abs/1603.05563}, \emph{Acta Astronaut.,} submitted, 2016.

\bibitem{paper9}
Janhunen, P., P.K.~Toivanen, J.~Polkko, S.~Merikallio, P.~Salminen,
E.~Haeggstr\"om, H.~Sepp\"anen, R.~Kurppa, J.~Ukkonen, S.~Kiprich,
G.~Thornell, H.~Kratz, L.~Richter, O.~Kr\"omer, R.~Rosta, M.~Noorma,
J.~Envall, S.~L\"att, G.~Mengali, A.A.~Quarta, H.~Koivisto,
O.~Tarvainen, T.~Kalvas, J.~Kauppinen, A.~Nuottaj\"arvi and
A.~Obraztsov, Electric solar wind sail: Towards test missions,
\emph{Rev.~Sci.~Instrum.,} \textbf{81}, 111301, 2010.

\bibitem{paper3}
Janhunen, P., On the feasibility of a negative polarity electric sail,
\emph{Ann.~Geophys.,} \textbf{27}, 1439--1447, 2009.

\bibitem{plasmabrake}
Janhunen, P., Electrostatic plasma brake for deorbiting a satellite,
\emph{J.~Prop.~Power,} \textbf{26}, 370--372, 2010.

\bibitem{paper24}
Janhunen, P., Simulation study of the plasma brake effect,
\emph{Ann.~Geophys.,} \textbf{32}, 1207--1216, 2014.

\bibitem{KestilaEtAl2013}
Kestil\"a, A., T.~Tikka, P.~Peitso, J.~Rantanen, A.~N\"asil\"a,
K.~Nordling, H.~Saari, R.~Vainio, P.~Janhunen, J.~Praks and M.~Hallikainen,
Aalto-1 nanosatellite -- technical description and
mission objectives. \emph{Geosci. Instrum. Method. Data
Syst.,} \textbf{2}, 121--130, 2013.

\bibitem{Koln2014}
Janhunen, P., Coulomb drag devices: electric solar wind sail
propulsion and ionospheric deorbiting, Space Propulsion 2014, K\"oln,
Germany, May 19-22, 2014 (\url{http://arxiv.org/abs/1404.7430}).

\bibitem{ASTRONUM2011}
Janhunen, P., PIC simulation of Electric Sail with explicit trapped
electron modelling, ASTRONUM-2011, Valencia, Spain, June 13--17, \emph{ASP
Conf. Ser.,} \textbf{459}, 271--276, 2012 (\url{http://www.electric-sailing.fi/papers/ASTRONUM2011.pdf}).

\bibitem{paper6}
Janhunen, P., Increased electric sail thrust through removal of
trapped shielding electrons by orbit chaotisation due to spacecraft
body, \emph{Ann.~Geophys.,} \textbf{27}, 3089--3100, 2009.

\bibitem{paper35}
Janhunen, P., Boltzmann electron PIC simulation of the E-sail effect,
\emph{Ann. Geophys.,} \textbf{33}, 1507--1512, 2015.

\bibitem{SiguierEtAl2013}
Siguier, J.-M., P.~Sarrailh, J.-F.~Roussel, V.~Inguimbert, G.~Murat
and J.~SanMartin, Drifting plasma collection by a positive biased
tether wire in LEO-like plasma conditions: current measurement and
plasma diagnostic, \emph{IEEE Trans.~Plasma Sci.,} \textbf{41}, 3380--3386, 2013.

\bibitem{LattEtAl2014}
L\"att, S., A.~Slavinskis, E.~Ilbis, U.~Kvell, K.~Voormansik, E.~Kulu,
ESTCube-1 nanosatellite for electric solar wind
sail in-orbit technology demonstration, \emph{Proc.\@ Estonian Acad.\@
  Sci.}, \textbf{63}(2S), 200--209, 2014.

\bibitem{EnvallEtAl2014}
Envall, J., P.~Janhunen, P.~Toivanen, M.~Pajusalu, E.~Ilbis, J.~Kalde,
M.~Averin, H.~Kuuste, K.~Laizans, V.~Allik, T.~Rauhala, H.~Sepp\"anen,
S.~Kiprich, J.~Ukkonen, E.~Haeggstr\"om, T.~Kalvas, O.~Tarvainen,
J.~Kauppinen, A.~Nuottaj\"arvi and H.~Koivisto, E-sail test payload of
ESTCube-1 nanosatellite, \emph{Proc.~Estonian Acad.~Sci.,},
\textbf{63}, 210--221, 2014.

\bibitem{SeppanenEtAl2011}
Sepp\"anen, H., S.~Kiprich, R.~Kurppa, P.~Janhunen and
E.~Haeggstr\"om, Wire-to-wire bonding of um-diameter aluminum wires
for the Electric Solar Wind Sail, \emph{Microelectronic Engineering,} \textbf{88},
3267--3269, 2011.

\bibitem{SeppanenEtAl2013}
Sepp\"anen, H., T.~Rauhala, S.~Kiprich, J.~Ukkonen, M.~Simonsson,
R.~Kurppa, P.~Janhunen and E.~Haeggstr\"om, One kilometer (1 km)
electric solar wind sail tether produced automatically,
\emph{Rev.~Sci.~Instrum.,} \textbf{84}, 095102, 2013.

\bibitem{KleshchEtAl2015}
Kleshch, V.I., E.A.~Smolnikova, A.S.~Orekhov, T.~Kalvas, O.~Tarvainen,
J.~Kauppinen, A.~Nuottaj\"arvi, H.~Koivisto, P.~Janhunen and
A.N.~Obraztsov, Nano-graphite cold cathodes for electric solar wind
sail, \emph{Carbon}, \textbf{81}, 132--136, 2015.

\bibitem{BilitzaEtAl2011}
Bilitza, D., L.-A.~McKinnell, B.~Reinisch and T.~Fuller-Rowell, The
International Reference Ionosphere (IRI) today and in the future,
\emph{J.~ Geodesy,} \textbf{85}, 909--920, 2011.























\end{thebibliography}
\end{document}